# Recent applicable delivery approaches of peptide nucleic acids to the target cells

Reyhane Alidousti[1], Mostafa Shakhsi-Niaee[2]

[1] Departments of Genetics, Faculty of Science, Shahrekord University, Shahrekord, Iran,
reyhanealidousti@ymail.com
[2] Departments of Genetics, Faculty of Science, Shahrekord University, Shahrekord, Iran, Iran,
shakhsi-niaei.m@sci.sku.ac.ir

**Abstract**
Peptide nucleic acids (PNAs) are artificial nucleic acids with a peptide backbone instead of sugar phosphate backbone of DNA or RNA. Their resistance to degradation, selectivity and greater binding affinity in comparison to usual nucleic acids led to consideration of their great potential for different applications. For example, they can be used in molecular diagnostics and antisense therapeutics such as antimicrobial agents or gene regulatory tools. On the other hand, large hydrophilic property of PNA molecules, which inhibit them to cross cell membranes readily, is an obstacle for their delivery to considered target cells and a limiting criterion for their applications. Therefore, PNA delivery technologies have been developing to hurdle this limitation. For example, addition of lysine residues, charged membrane penetrating peptide sequences, PNAs conjugated with antibodies or steroids, cationic liposomes as carriers of PNA conjugates, protective peptides and technology of photochemical internalization (PCI) as well as by the recent technology of nanoparticle-based delivery have been employed. In this article we compared different delivery technologies which can be applicable to PNAs. As a result nanoparticle-based delivery showed more advantages in comparison to others and its application is growing fast.

**Keywords:** Target cell delivery, Peptide nucleic acid, PNA, Delivery.

## 1. Introduction

Peptide nucleic acid (PNAs) are artificial nucleic acid with unique physicochemical properties which come from their peptide like *N*-(2-aminoethyl)-glycine backbone instead of the sugar phosphate backbone of natural nucleic acids [1].
Peptide nucleic acids (PNAs) are synthetic structural analogues of DNA and RNA. They recognize specific cellular nucleic acid sequences and form stable complexes with complementary DNA or RNA [2].
PNAs mimic DNA probes that bind with a high sequence specificity in binding to complementary strands of DNA and/or RNA and therefore can interfere with expression of targeted genes [3]. In addition, PNAs have higher chemical stability than natural nucleic acids. PNA: DNA and PNA: RNA duplexes show increased thermodynamic stability when compared with DNA: DNA and DNA: RNA duplexes. This property is partially because the uncharged PNA backbone and negatively charged DNA or RNA backbone can repulse each other less than other DNA or RNA combinations [4]. Due to these features, PNAs have an exceptional potential for therapeutic applications and diagnostic use [5].
Using PNAs is a promising approach for manipulation of mitochondrial DNA (mtDNA) replication and expression within mammalian cells as a not easily available DNA material [6].
Therefore, PNAs can be used for silencing genes especially those are critical for bacterial viability or growth, with an antibacterial application [3, 7]. This ability besides Increasing number of pathogens that are resistant to usual therapeutics on one hand, and the difficulties of new antimicrobials recognition on the other hand have prompted focusing on new approaches such as antisense PNA [8]. PNA shows also great potential for use in molecular diagnostics and antisense therapeutics [9] which is because of its resistance to degradation by nucleases and proteases [10], greater binding affinity, specificity [4] and strand invasion capability in comparison to usual nucleic acids [11]. Strand invasion behavior of the neutral PNA backbone provides a potential for detection of duplex DNA and put the antigene PNAs (agPNAs) as an applied tool for investigation of the structure and function of chromosomal DNA [12].
However, in the first place benefit from advantages of PNAs requires straightforward strategies for their delivery into cells.

### 1, 1. Applicable delivery technologies for PNAs

Despite these positive gene-targeting attributes, unmodified PNA resists cellular uptake and therefore has limited bioavailability [13]. This limitation has impeded its development as a candidate therapeutic, although efficiency of cellular delivery of PNA in vitro has been achieved in some extent by the addition of lysine residues [14], charged membrane penetrating peptide sequences [13], PNAs conjugated with antibodies or steroids [15, 13], cationic liposomes as carriers of PNA conjugates [16] and protective peptides [17]. Peptide nucleic acids can also enter and release into the cell by the novel technology of photochemical internalization (PCI) [18] or by the recent technology of nanoparticle-based delivery [19].

### 1,2. Delivery of negatively charged PNA

PNAs show large hydrophilic property [20] and therefore do not readily cross cell membranes [21]. As most of nucleic delivery technologies based on the negatively charged backbone of usual nucleic acids, it seems that by impart of some negative charge to PNA without disturbing its binding affinity with DNA or RNA, not only we are able to take advantage of these kind of delivery systems but also benefit of PNA with enhanced *in vivo* efficacy. For example PNAs were synthesized conjugated to oligophosphonates via phosphonate glutamine and bisphosphonate lysine amino acid derivatives thereby introducing up to twelve phosphonate moieties into a PNA oligomer. It is reported that this modification of the PNA does not interfere with the nucleic acid target binding affinity based on thermal stability of the PNA/RNA duplexes. For confirmation of its functionality it was delivered to cultured HeLa pLuc705 cells by Lipofectamine, the PNAs showed dose-dependent nuclear antisense activity in the nanomolar range as inferred from induced luciferase activity as a consequence of pre-mRNA splicing correction by the antisense-PNA [22].

In a newer research an oligoaspartic acid–PNA conjugate was designed and its delivery investigated by using the conventional cationic carriers, such as polyethylenimine (PEI) and Lipofectamine 2000. Enhanced delivery into cells achieved by monitoring its splicing correction efficiency. The negatively charged oligo-aspartic acid–PNA (Asp (n)–PNA) formed complexes with PEI and Lipofectamine, and the resulting Asp (n)–PNA/PEI and Asp (n)–PNA/Lipofectamine complexes were introduced into cells [2].

### 1,3. Delivery of PNA using cationic carriers

In a research cationic liposomes were used to efficiently complexate DNA–PNA hybrid molecules and mediate their binding to target cells. PNAs are not suitable for an efficient delivery with commonly used liposomal formulations but transfection of PNA–DNA hybrid molecules to in vitro cultured cells could be of great interest to determine the applications of these new reagents to experimental alteration of gene expression [23].

In a different study cationic submicron particles constituted with Eudragit RS 100, plus different cationic surfactants, such as dioctadecyl-dimethyl-ammonium bromide (DDAB18) and diisobutylphenoxyethyl-dimethylbenzyl ammonium chloride (DEBDA), was produced. This cationic complex used as a transport and delivery system for PNA-DNA chimeras as well as DNA/DNA and DNA/ PNA hybrids. Submicron particles could offer advantages over other delivery systems because they maintain unaltered physicochemical properties for long time periods, allowing long-term storage, and are suitable for industrial production [24].

Another group of researchers studied an intracellular biodegradable triphenylphosphonium (TPP) cation based transporter system. In this system, TPP is linked, via a biolabile disulfide bridge, to an activated mercaptoethoxycarbonyl moiety, allowing its direct coupling to the N-terminal extremity of a free PNA through a carbamate bond [25].

### 1,4. Conjuvation of polyaromatic, lipophilic compounds to PNA

It is also reported that conjugation of (hetero) polyaromatic, lipophilic compounds such as 9-aminoacridine, benzimidazoles, carbazole, anthraquinone, porphyrine, psoralen, pyrene, and phenyl-bis-benzimidazole ("Hoechst") to PNA can dramatically improve liposome-mediated cellular delivery both to cytoplasm as well as to the nucleus [26].

### 1,5. Delivery liposomal system

Cationic lipids have been widely used for the delivery of charged oligonucleotide (ON) analogues but most of the commercial formulations are toxic and poorly stable in the presence of serum proteins. Therefore, DOGS/DOPE liposome formulation named DLS (for delivery liposomal system), have developed that allows for the efficient nuclear delivery of negatively charged antisense ON analogues as monitored by fluorescence microscopy and by their ability to correct deficient pre-mRNA splicing, even in serum-supplemented cell culture [27]. A new liposomal formulation composed of egg PC/cholesterol/DSPE-PEG2000 was successfully loaded with PNA or fluorescent PNA oligomers. In this research PNA loaded liposomes were able to efficiently and quickly promote the delivery of a PNA which targeted the microRNA miR-210 in cell culture. This methodcould down-regulate miR-210 at a low concentration of considered PNA [28].

### 1,6.DNA/lipid complex Delivery

In some experiments a complex of complementary DNA oligonucleotides, cationic lipid and PNAs were delivered into cultured human cells [29]. This method is a variation of standard protocols for lipid–mediated transfection. The DNA binds to the PNA, the lipid binds to the DNA, and the PNA is transported into cells as cargo by the DNA/lipid complex [12].

PNAs can also be complexed with a conjugate of a lipid domain (fatty acid) and a cationic peptide (called a CatLip conjugate). This complex can increase the biological effect of the corresponding PNA up to 2 orders of magnitude. In this assay the length of the fatty acid (C8−C16) is correlated with effect of PNA as well as increased cellular toxicity. It is reported that the relative progress is significantly higher for Tat peptide compared to oligoarginine. Chloroquine treatment can help lipophilic domain to increase the endosomal uptake as well as endosomal escape which leads to improving the bioavailability of even larger hydrophilic molecules [13].

**1,7. Cell Penetration Peptides**

The cell membrane is the structure that protects living cells from the surrounding environment, only allowing the movement of compounds generally with small molecular size across this barrier into the cell. A variety of methods have been developed to improve PNA uptake into cells, and the currently favored approach involves conjugation to cell penetrating peptides (CPPs) [30].

CPPs also have been called by different names such as protein translocation domain, membrane translocating sequence, Trojan peptide. CPPs are rich in basic amino acids such as arginine and lysine and are able to translocate over membranes and gain access to the cell interior. They can deliver large-cargo molecules, such as oligonucleotides, into cells [31].

Generally, CPPs are defined as short, water-soluble and partly hydrophobic, and/or polybasic peptides (at most 30–35 amino acids residues) with a net positive charge at physiological pH [32]. The main feature of CPPs is that they are able to penetrate the cell membrane at low micromolar concentrations in vivo and in vitro without using any chiral receptors [33]. CPPs can be divided into cationic, amphipathic and hydrophobic groups. CPPs may use different mechanisms such as direct penetration, endocytosis or by formation of a transitory structure [34]. Furthermore, and even more importantly, these peptides are capable of internalizing electrostatically or covalently bound biologically active cargoes with high efficiency and low toxicity [32, 33]. For example, they are able to transport small molecules such as plasmid DNA, small interfering RNA, as well as proteins, viruses and other nanoparticles across the cell membrane without any significant damages to cell membrane [34].

CPPs are categorized into the different subgroups based on their individual properties. One of the classifications is based on the origin of the peptide. It includes protein-derived peptides such as TAT and penetratin, which are also called protein transduction domains (PTDs). The second subgroup is the chimeric peptides which may contain two or more motifs from other peptides, for instance, transportan derived from mastoparan and galanin and its shorter analogue TP10. Synthetic peptides are another group in this category such as the polyarginine family [33].

In another classification based upon different peptide sequences and binding properties to the cell membrane lipids CPPs are divided into three classes. These classes include primary amphipathic, secondary amphipathic and nonamphipathic CPPs [35]. Primary amphipathic CPPs (paCPPs), such as transportan [36] or TP10 [37] contain typically more than 20 amino acids. They have sequentially hydrophobic and hydrophilic residues along their primary structure [35]. In addition to endocytosis, the proposed mechanism for this group of CPPs is direct membrane transduction. Model studies have suggested that the direct transduction occurs via pore formation, carpet-like perturbations, or inverted micelles formed in the bilayer membrane [38]. Some primary amphipathic CPPs such as TP10 are toxic to cells even at low concentrations. In addition, amphipathic CPPs interact with both natural and anionic lipid membranes [35]. Secondary amphipathic CPPs (saCPPs), such as penetratin [39], pVEC [40], and M918 [41] often contain a smaller number of amino acids compared with primary amphipathic CPPs. Their amphipathic property is revealed when they form an alpha-helix or a beta sheet structure upon interaction with a phospholipid membrane. They typically bind to model membranes with a certain fraction of anionic lipids [35].

The third class, that is, the nonamphipathic peptides (naCPPs) are rather short with a high content of cationic amino acids (arginine) such as R9 [42] and TAT(48–60). They bind to the lipid membrane with a high amount of anionic lipids. Membrane leakage is not observed at low micromolar concentrations [31]. naCPPs and saCPPs are both less toxic than paCPPs, and higher concentrations or application of a transmembrane potential seems to be required to make the membrane unstable, both in the cell and in membrane model systems. It has been shown that acylation of these cationic peptides to make them more hydrophobic is a way to induce membrane leakage by this class of CPPs [43].

Cationic CPPs have been used largely for intracellular delivery of low-molecular-mass drugs, biomolecules and particles. Most cationic CPPs bind to cell-associated glycosaminoglycans and endocytosis is the main way of internalisation [44]. However, release from endosomes into cytosol remains a rate-limiting step for their delivery system. Therefore, some scientists described protocols for the delivery of CPP-PNA conjugates into cells by different treatments such as photochemical internalization, Ca2+, or chloroquine treatment [45]. For example, conjugation of an arginine-rich CPPs to PNA allows efficient nuclear delivery in the absence of chloroquine as monitored in a splicing correction assay [27] orcombining CPP-PNA conjugate administration with a photochemical internalization technique using photosensitizers such as aluminum phthalocyanine (AlPcS(2a)) or tetraphenylporphyrin tetrasulfonic acid (TPPS) was a successful method for overwhelming this problem [46]. Some studies have been attempted by incorporating positively charged residues such as lysine and arginine [15] or arginine and tryptophans [21] to the PNA molecules to enhance the PNA delivery efficiency, especially in the presence of the endosomolytic agent chloroquine [21]. However, Cordier et al., 2014 showed that Replacement of arginines with lysines lead to about six-fold reduction in the entery of the conjugate, emphasizing crucial role of arginines for conjugate uptake.

### 1, 8. Electroporation

There is successful introduction of a PNA by electroporation to human prostate cancer cells grown in culture which could target the coding region of *IGF-1R* mRNA and inhibited translation elongation [47].

### 1, 9. Conjugation of PNAs to a Protective antigen

Wright et al (2008) employed a non-toxic component of anthrax toxin as a protective antigen (PA) for PNA delivery to the cell. They revealed that anthrax PA could efficiently transport PNA into cells without conferring its functionality in comparison with PNA 18mers with C-terminal poly-lysine tails [17]. Marlin et al., 2012 confirmed that conjugation of PNAs to flavin (an isoalloxazine) can lead to efficient internalization into cells through an endocytic pathway. In their report, endosomal release of flavin-conjugates into the cell increased by chloroquine treatment [48].

### 1, 10. Conjugation of PNAs to a nanoparticles

Nanoparticle-based delivery methods has been recently developed because of lack of specificity and in many cases the problem of carriers toxicity. This method has the potential of targeted cell delivery and controled delivery of PNA-based therapeutics. For example the PLGA and PBAE: PLGA nanoparticles have been used to deliver PNAs into the cytoplasm and nucleus of cells, where antisense interference by PNAs are required. As PLGA and PBAE: PLGA nanoparticles were not toxic in mice. Some other materials such as zeolite and mesoporous silica has been suggested for encapsulation of PNAs. Therefore, engineered nanocarriers conjugated with targeting ligands may considered for diagnostic and therapeutic applications of PNAs [19].

### Conclusion

Peptide nucleic acids (PNAs) are considered as a powerful tool to regulate target gene expression. However, PNAs can not penetrate through the cell membrane similar most of the DNA and RNA, for example because of their uncharged property. Therefore, different strategies have been considered to enhance their delivery efficiency. However, Nanoparticle-based delivery methods seem better strategy because of their lack of toxicity for the cell as well as potential of target cell therapy.